\documentclass[final]{aipproc}

\newcommand{\be}{\begin{equation}} \newcommand{\ee}{\end{equation}}
\newcommand{\bea}{\begin{eqnarray}} \newcommand{\eea}{\end{eqnarray}}
\newcommand{\bean}{\begin{eqnarray*}}
  \newcommand{\eean}{\end{eqnarray*}}

 \newcommand{\gapproxeq}{\lower
  .7ex\hbox{$\;\stackrel{\textstyle >}{\sim}\;$}}
\newcommand{\lapproxeq}{\lower .7ex\hbox{$\;\stackrel{\textstyle
      <}{\sim}\;$}}

 \newcommand{\bc}{\begin{center}}
  \newcommand{\ec}{\end{center}} \newcommand{\btab}{\begin{tabular}}
  \newcommand{\etab}{\end{tabular}} 
  \def\qq{$ q\bar q $} \def\ss{$ s\bar s $}
 \def\uu{$u\bar u$} \def\dd{$d\bar d$}
\def\cc{$c\bar c$} \def\bb{$b\bar b$}
\def\10bar{$\bar {\hbox{\bf 10}}$} \def\3bar{$\bar {\hbox{\bf 3}}$}
\layoutstyle{6x9}

\begin{document} 
 
\title{The End of the Constituent Quark Model?}  \author{F.E.
  Close}{address={Department of Theoretical Physics, University of
    Oxford, \\ 1 Keble Rd., Oxford, OX1 3NP, United Kingdom; \\ 
    f.close@physics.ox.ac.uk}}
 
\begin{abstract}  
  In this conference summary talk at Hadron03, questions and
  challenges for Hadron physics of light flavours are outlined.
  Precision data and recent discoveries are at last exposing the
  limitations of the naive constituent quark model and also giving
  hints as to its extension into a more mature description of hadrons.
  These notes also pay special attention to the positive strangeness
  baryon $\Theta^+(1540)$ and include a pedagogic discussion of
  wavefunctions in the pentaquark picture, their relation with the
  Skyrme model and related issues of phenomenology.
\end{abstract} 
\maketitle

\subsection*{Introduction} 

My brief was to concentrate on light hadrons; but where do heavy
hadrons end and light begin?  I shall focus on what heavy flavours can
teach us about light, and vice versa.  The possible discovery of an
exotic and metastable baryon with positive strangeness, the
$\Theta^+(1540)$, has led to an explosion of interest in recent months
and throughout this conference. There was a dedicated discussion
session about it, which highlighted much confusion. In the hope of
clarifying some of the issues, I have decided to devote a considerable
part of this summary to a pedagogic description of wavefunctions and a
review of some of the emerging literature that drew comment at the
conference.

\subsection*{Light Hadron Spectroscopy and Dynamics: Present and Future}

As regards the future of light hadrons experimentally: we have heard
of several examples of innovative methods involving high energy
machines. In particular, we have electron positron machines designed
as B-factories, which turn out to access lower energies as a result of
the initial state radiation\cite{isr}. They also provide copious data
on $\gamma \gamma \to $ light hadrons. At HERA we have vector mesons
produced diffractively but also now charge conjugation positive states
produced apparently in the rapidity gap between the photon and the
target proton\cite{heracentral}.  Then there is the new opportunity
for central production in proton-proton collisions at STAR and the
proposal for this at HERA-g \cite{schlein}. Exploiting the $dk_T$
filter and $\phi$ dependences may separate \qq~ states from those with
gluonic admixtures or S-wave substructure (see later)\cite{ck}.

Then we have heard about B and D decays into light hadrons from BaBar,
BELLE and FOCUS and novel ideas from Ochs about $B \to K^* + 0^{++}$
as an entree into the light scalars\cite{ochs}. There is also $D_s \to
\pi (s\bar{s} )_{0^{++}}$ which gives an entree to the scalar \ss~
sector.  The decays $\psi \to \gamma \gamma V$, where $V \equiv \rho,
\phi$ have been reported from BES\cite{bes} and will be pursued by
CLEO-c \cite{cleoc}. So far these data have been applied to the
$\iota(1440)$\cite{bes} region but eventually promise to provide
information on the flavour content of any $C=+$ mesons, through $\psi
\to \gamma R(C=+) \to \gamma \gamma V$. In particular for $R \equiv
0^{++}$ this can give essential information on the flavour contents,
and hence mixings with the glueball of lattice QCD, of the various
scalar mesons\cite{donnachie}.
 
When high statistics data are available at CLEO-c and BES-III we can
study $\chi \to \pi + R$ where $R \equiv$ light hadrons and complement
the old $p\bar{p}$ data from LEAR. The bonus will be that $\sqrt{s}
\sim 3.5$GeV and that the overall $J^{PC}$ is known in the subsequent
partial wave analyses.  Of particular interest here could be $\chi
\equiv 1^{++}$ where in S-wave the recoil system $R \equiv 1^{-+}$,
which is the exotic channel favoured for hybrid mesons. So there are
reasons to be optimistic about sorting out light hadron spectroscopy
and dynamics and solving whether and how the gluonic degrees of
freedom are manifested in the strong QCD regime.
 
As G.W.Bush and T. Bliar might summarise the search for missing
glueballs and hybrids: we know they exist; they are hidden but we will
find them; give us time - we have only been searching for 20 years.
 
We have also heard\cite{bes} how $\psi$ decays can give novel insights
into baryon resonances in the timelike region through $\psi \to
\bar{N} N^*$ or $\bar{\Delta} \Delta^*$. This selects isospin states
apart from a background due to the intermediate $\psi \to \gamma^*$
channel, and gives complementary information to that from the maturing
data from Jefferson Laboratory\cite{weygand}. Finally we have the
degeneracy of the hybrid candidate $\pi(1800)$ and $D(1865)$. The
Cabibbo suppressed decays of the latter\cite{msratcliffe} share common
channels with the strong decays of the former. Disentangling this is a
significant overlap between heavy and light flavours of a pragmatic
nature let alone the interesting potential implications for novel
physics.
 
A problem in light hadron dynamics is that not all of the data can be
correct. Swanson\cite{swanson} has shown a nice figure listing all of
the mesons as a function of $J^{PC}$ from the PDG which is clearly
overpopulated. This leads me to two requests: one to theorists and one
to experimentalists.
 
Theorists: beware of taking your favourite random model; finding the
$J^{PC}$ states that agree with it and then ignoring, excusing or
tweaking the model to apologise for those that do not not. It is
important to keep the big picture in mind if we are to progress. Focus
on the wood not the trees. There is information on more than
spectroscopy. We have decays and also production dynamics that can
provide essential constraints on models and interpretation.
 
Experimentalists: what am I supposed to regard as ``official" data? Is
a conference presentation, which does not get refereed for a peer
reviewed journal ``official"? At this Hadron series over the years we
have seen reports of measurements on say phenomenon H1.  Nothing is
seen in a peer reviewed journal. Two years later the same group might
report the phenomenon as H2; or they might say nothing and only when
questioned by those in the know reveal that they no longer see any
signal. However, this is not reported as an ``official" withdrawal.
It is hard enough trying to interpret the data without the added
pollution of work in process. In particular, it is extremely important
that states which are claimed, and then go away, be reported as
withdrawn.
 
I would not want to stifle the presentation of preliminary data, as
such creates discussion that can be mutually informative. However,
when it is written up for the proceedings I would urge that a clear
statement be made up front as to the status of the report: e.g. on
what timescale will a version be prepared for ``official" publication?
If the data are even more preliminary, I would suggest that no written
summary be produced for the proceedings; or that a clear disclaimer be
made that this is a report on an individual's analysis. I would hope
that any such presentation has received the endorsement of the group,
but suspect that this is not always the case, since at this conference
I have heard at least one parallel session where group members
appeared to be hearing of some analysis for the first time. This is
fine for enabling the ``critical filtering" that produces the best
analyses, but dangerous nonetheless if associated ``health warnings"
are not prominent.
 
As regards the (scalar) glueball and hybrid states it is time to move
on from simple ideas that such states exist in some pure sense. We
have heard many times here statements on the line of ``The $f_0(xxxx)$
is the scalar glueball", where you are invited to insert your number
of choice out of 970,1300,1500 or 1700. And for the hybrid: ``The
$\pi_1(yyyy)$ is exotic, ergo it is hybrid" (where $yyyy$ is 1400,
1600 or 1800). The only pattern seems to be that the former set
involve odd and the latter even numbers in their first two places.

What I offer here is not a solution, but needs to be taken into
account when seeking the solution.  The real world contains thresholds
for hadronic channels with the same $J^{PC}$ as these objects and will
involve mixings with those as well as between the primitive glueball
and $q\bar{q}$ flavoured states. So the scalar mesons with I=0 in the
PDG will be mixtures of glueball and flavoured $q\bar{q}$ at least.
(Hence the interest in $\psi \to \gamma \gamma V$ alluded to earlier
to help disentangle this mixing). Likewise with the $\pi_1$ states.
The non-relativistic quark model was built, in part, on the absence of
such exotic $J^{PC}$ combinations. Now we have three being claimed.
This is too much of a good thing and the presence of S-wave thresholds
such as $\pi b_1$ and $\pi f_1$ around 1400MeV surely plays some
essential role. Another school of thought has been presented here:
could some of the $\pi_1$ states be evidence for $qq\bar{q}\bar{q}$ in
$10 \pm $\10bar configurations\cite{chung}? Possibly, but beware the
dog that didn't bark in the night: invoking multiquarks to accommodate
one or two awkward states also implies the existence of whole
multiplets of associated states. The failure to see them also needs to
be explained in such models.

There is rather general agreement now that qualitatively the scalar
mesons sector contains a scalar glueball degree of freedom\cite{glue}
in the data, the question now is to quantify it. In this regard there
seem to be two broad schools\cite{glueballs1,glueballs2} and data need
to be able to distinguish between these as a minimum before we can
claim the glueball as proven. First their common features: there is a
scalar glueball present in the mass region up to around 1700MeV, which
mixes with and disturbs the ``simple" isoscalar $q\bar{q}$ sector.
Now for their details.  One\cite{glueballs1} is that the mesons above
1GeV, $f_0(1700;1500;1370)$ are the I=0 states of \qq~ mixed with the
$G$, and that $K(1430)$ and $a_0(1450)$ are the other members of the
extended nonet; in this scenario the mesons below 1GeV, in particular
the $f_0(980)$ and $a_0(980)$ have a $qq\bar{q}\bar{q}$ or dimeson
dynamical structure\cite{jaffe,tornqvist}. The other\cite{glueballs2}
is that the $f_0(980)$ and $a_0(980)$ are in the nonet with the
$f_0(1500)$ and $K(1430)$ (the $f_0(980)$ and $f_0(1500)$ having an
interesting ``inversion" of properties in $SU(3)$ flavour to the
pseudoscalar $\eta$ and $\eta'$); the $f_0(1370)$ is not recognised as
a real resonance state, the $a_0(1450)$ if it exists is in some other
nonet (perhaps with the $f_0(1700)$; the $\kappa$ is not resonant and
the $\sigma(600)$ is part of a very broad scalar glueball whose
effects are felt throughout an extended energy range.
 
As I have shares in one of the above pictures, the following
observations on the novel meson states reported at this conference
might be distorted by my prejudices, but with that caveat in mind I
offer them for consideration nonetheless.
 
 \subsection*{When does the quark model work?}
 
 \vskip 0.1in

 There is general agreement that the NRQM is a good phenomenology for
 \bb~ and \cc~ states below their respective flavour thresholds.
 Taking \cc~ as example we have S states ($\eta_c,\psi, \psi(2S)$),
 P-states ($\chi_{0,1,2}$) and a D-state ($\psi(3772)$), the latter
 just above the $D\bar{D}$ threshold. Their masses and the strengths
 of the E1 radiative transitions between $\psi(2S)$ and $\chi_J$ are
 in reasonable accord with their potential model status. In particular
 there is nothing untoward about the scalar states.
 
 Do the same for the light flavours and one finds clear multiplets for
 the $2^{++}$ and $1^{++}$ states (though the $a_1$ is rather messy);
 it is when one comes to the scalars that suddenly there is an excess
 of states. An optimist might suggest that this is the first evidence
 that there is an extra degree of (gluonic) freedom at work in the
 light scalar sector. But there is more: there is a clear evidence of
 states that match onto either $qq\bar{q}\bar{q}$ or correspondingly
 meson-meson in S-wave.
 
 Such a situation is predicted by the attractive colour-flavour
 correlations in QCD\cite{jaffe,tornqvist}. Establishing this has
 interest in its own right but it is also necessary to ensure that one
 can classify the scalar states and then identify the role of any
 glueball by any residual distortion in the spectrum. It is in this
 context that discoveries this year of narrow states in the heavy
 flavour sector provide tantalising hints of this underlying dynamics
 elsewhere in spectroscopy. If this is established it could lead to a
 more unified and mature picture of hadron spectroscopy.
 
 The sharpest discoveries this year have involved narrow resonances:
 $c\bar{s}$ states, probably $0^+,1^+$, lying just below $DK,D^*K$
 thresholds; and \cc degenerate with the $D^oD^{*o}$ threshold. These
 are superficially heavy flavour states and out of my remit, but their
 attraction to these thresholds involves light quarks and links to a
 more general theme which I shall develop below.
 
 First note how we have been reminded here by Barnes\cite{barnes} that
 the \cc~ potential picture gets significant distortions from the $DD$
 threshold region, such that even the \cc~ $\chi$ states can have 10\%
 or more admixtures of meson pairs, or four quark states, in their
 wavefunctions.  Also Cahn\cite{cahn} reminded us that the simple
 potential models of the $D_s$ states are inadequate to explain the
 2.32GeV and 2.46GeV masses of these novel states as simply $c\bar{s}$
 in some potential.  Furthermore, Davies\cite{davies} showed that the
 lattice seems to prefer the masses to be higher than actually
 observed, though the errors here are still large. In summary there is
 an emerging picture that these data on the $D_s$ sector (potentially
 $0^+$ and $1^+$ and the S-wave $DK$ and $D^*K$ thresholds) and the
 \cc~ sector (with the S-wave $D^oD^{*o}$ thresholds) confirm the
 suspicion that the simple potential models fail in the presence of
 S-wave continuum threshold(s).
 
 Now let's examine this in the light flavoured sector. The multiplets
 where the quark model works best are those where the partial wave of
 the \qq~ or $qqq$ is lower than that of the hadronic channels into
 which they can decay. For example, the $\rho$ is S-wave \qq~ but
 P-wave in $\pi\pi$; as S-wave is lower than P-wave, the quark model
 wins; by contrast the $\sigma$ is P-wave in \qq~ but S-wave in
 $\pi\pi$ and in this case it is the meson sector that wins and the
 quark model is obscured.
 
 A similar message comes from the baryons. The quark model does well
 for the $\Delta$ (S-wave in $qqq$ but P-wave in $\pi N$); at the
 P-wave $qqq$ level it does well for the $D_{13}$, which as its name
 implies is D-wave in hadrons, but poorly for the $S_{11}$ which is
 S-wave in $N\eta$.  The story repeats in the strange sector where the
 strange baryons with negative parity would be $qqq$ in P-wave: the
 $D_{03}(1520)$ is fine but the $S_{01}(1405)$ is the one that seems
 to be contaminated with possible $KN$ bound state effects.
 
 As an exercise I invite you to check this out. It suggests a novel
 way of classifying the Fock states of hadrons. Instead of classifying
 by the number of constituent quarks, list by the partial waves with
 the lowest partial waves leading. Thus for example
 
 \[
 0^{++} = |0^-0^- (qq\bar{q}\bar{q}) \rangle_S + |q\bar{q} \rangle_P
 +....
 \]
 
 while
 \[  
 1^{--} = |q\bar{q} \rangle_S + |0^-0^- (qq\bar{q}\bar{q}) \rangle_P +
 ....
 \]
 or
 \[  
 \Delta(1230) = |qqq \rangle_S + |\pi N (qqqq\bar{q})\rangle_P +...
 \]
 This holds true for
  \[
  2^{++} = |q\bar{q} \rangle_P + |0^-0^- (qq\bar{q}\bar{q}) \rangle_D
  +....
 \]
 the relevant S-wave vector-meson pairs being below threshold. For the
 remaining P-wave \qq~ nonet with C=+ we have a delicate balance
  \[
  1^{++} = |q\bar{q} \rangle_P + |0^-1^-
  \rangle_S + ....
 \]
 where the $\pi \rho$ S-wave distorts the \qq~ $a_1$, as is well
 known; the $f_1(1285)$ is protected because the two body modes are
 forbidden by G-parity; for the strange mesons the $K^*\pi$ and $K
 \rho$ channels play significant roles in mixing the $1^{++}$ and
 $1^{+-}$ states while the \ss~ state is on the borderline of the
 $KK^*$ threshold.
 
 Chiral models which focus on the hadronic color singlet degrees of
 freedom are thus the leading effect for the $0^{++}$ sector but
 subleading for the vectors.  An example was presented\cite{palaez}
 where the $N_c$ dependence of the coefficients of the chiral
 Lagrangian was studied. In the $N_c \to \infty$ limit it was found
 that $\Gamma(\rho) \to 0$, like \qq~ whereas $\Gamma(\sigma) \to
 \infty$, like a meson S-wave continuum. Thus there appears to be a
 consistency with the large $N_c$ limit selecting out the leading
 S-wave components.
 
 Conversely, the ``valence" quark model can give the leading
 description for the vectors or the $\Delta$ but there will be
 corrections that can be exposed by fine detail data. The latter are
 now becoming available for the baryons from Jefferson Laboratory; the
 elastic form factors of the proton and neutron show their charge and
 magnetic distributions to be rather subtle, and the transition to the
 $\Delta$ is more than simply the M1 dominance of the quark model.
 There are E2 and scalar multipole transitions which are absent in the
 leading $qqq$ picture. The role of the $\pi N$ cloud is being
 exposed; it is the non-leading effect in the above classification
 scheme. As we shall see later, the $\Theta^+(1540)$ as a pentaquark
 inspires novel insights into a potential pentaquark - or $N\pi$ cloud
 - component in the $N$ and $\Delta$.
 
 The message is to start with the best approximation - quark model or
 chiral - as appropriate and then seek corrections.
 
 Bearing these thoughts in mind it highlights the dangers of relying
 too literally on the quark model as a leading description for high
 mass states unless they have high $J^{PC}$ values for which the
 S-wave hadron channels may be below threshold. It also has
 implications for identifying hadrons where the gluonic degrees of
 freedom play an explicit role and cannot simply be subsumed into the
 collective quasi-particle known as the constituent quark.  Such
 states are known as glueballs and hybrids.
 
 The lightest glueball is predicted to be scalar\cite{glue} for which
 the problems arising from the S-wave thresholds have already been
 highlighted. At least here, by exploiting the experimental strategies
 outlined at the start of this talk, we are possibly going to be able
 to disentangle the complete picture. For the $2^{++}$ and $0^{-+}$
 glueballs above 2 GeV there are copious S-wave channels open, which
 will obscure the deeper ``parton" structure. Little serious thinking
 seems to have been done here.
 
 For the exotic hybrid nonet $1^{-+}$ we have a subtlety. In the
 flux-tube models abstracted from lattice QCD, the \qq~ are in an
 effective P-wave\cite{ip95,cd03}, which we may describe by
 $|q\bar{q}g \rangle_P $.  There is a leading S-wave $0^- 1^+$ meson
 pair at relatively low energies, such that

 \[
 1^{-+} = |0^-1^+ \rangle_S + |q\bar{q}g \rangle_P +....
 \]
 The S-wave thresholds for $\pi b_1$ and $\pi f_1$ are around 1400MeV,
 which is significantly below the predicted 1.8GeV for lattice or
 model hybrids and tantalisingly in line with one of the claimed
 signals for activity in the $1^{-+}$ partial wave. All is not lost
 however; a \qq~ or \qq$g$ nonet will have a mass pattern and decay
 channels into a variety of final states controlled by Clebsch-Gordan
 coefficients whereas thresholds involve specific meson channels.
 These can in principle be sorted out, given enough data in a variety
 of production and decay channels, but it may be hard.

\section*{Skyrmion meets the Quark}
 
In the above we have discussed where components beyond the leading
\qq~ or $qqq$ may obscure the simple quark model. We now come to a
case where the {\bf leading} component involves five constituents. If
this discovery is confirmed it will make a sobering reminder that
there can be phenomena latent in data that have been overlooked
perhaps for decades.

In the textbooks, one of the major planks in establishing the
constituent quark model is the absence of baryons with strangeness +1.
The announcement of such a particle, and with a narrow width is
therefore startling, if confirmed\cite{leps}. It is easy to
accommodate positive strangeness; you just allow an extra \qq~ to be
present, e.g. $uudd\bar{s}$. The problem though is that such a state
would be expected to fall apart so rapidly that its width would be
broad.  A narrow width, signaling metastability, therefore implies the
existence of some inhibiting factor. Its parity is undetermined and
that could itself discriminate among models. The state was predicted
in the Skyrme model\cite{skyrme} where it is a member of a \10bar with
$J^P = \frac{1}{2}^+$. This is already an interesting conundrum for a
quark model where the naive expectation is that the lightest state of
a pentaquark $uudd\bar{s}$ has all constituents in a relative S-wave,
hence $J^P = \frac{1}{2}^-$. However, this is true only when all the
quarks are treated symmetrically. There is a considerable literature
that recognises that $ud$ in colour $\bar{3}$ with net spin 0 feel a
strong attraction, which might even cause the S-wave combination to
cluster as $[udu][d\bar{s}]$ which is the S-wave KN system, while the
P-wave positive parity exhibits a metastability such as seen for the
$\Theta$. Two particular ways of realising this are due to Karliner
and Lipkin\cite{kl1} and Jaffe and Wilczek\cite{jw}, which I will come
to shortly.

A challenge for all quark models is the metastability of the state.
The historical stability of the strange hadrons was due to their
strong decays being forbidden; the first attempt to describe the
$\Theta$ as a pentaquark\cite{caps} built on this idea by proposing
that $\Theta$ be an isotensor resonance with states ranging from
$uuuu\bar{s}$ with charge +3 to $dddd\bar{s}$ with charge -1. This can
give a narrow width as there is no simple decay path that preserves
isospin.  The colour structure of such a pentaquark system is well
defined. The I=2 flavour-space is totally symmetric and so is totally
antisymmetric in colour-spin. This forces either $6_c \times (S=0)$ or
$3_c \times (S=1)$. Only the latter can combine with the $\bar{s}$ in
$\bar{3}_c$ to make the colour-singlet baryon. This leads to overall
$J^P = \frac{1}{2}^-$ or $ \frac{3}{2}^-$.  The price, or excitement,
is that there is a multiplet of states ($\Theta^{+++}....\Theta^-$) to
be found. This may be already ruled out if the ELSA\cite{leps} data
are confirmed as they find the $\Theta^+$ but have no evidence for any
partner, and thus suggest it is I=0.

Models with I=0 suggest that it is at the pinnacle of a flavour
\10bar, which is where the original Skyrme prediction would place it.
Thus there is also the interesting question of whether or under what
circumstances there is any correspondence between the Skyrme and quark
pictures.

Attempts to describe this as a pentaquark have been criticised in some
quarters on the lines that it is meaningless to describe a hadron as
made from a fixed number of quarks or antiquarks.  Let's first make
some obvious pedagogic remarks in order to accommodate some
suggestions that I shall make later.

When the proton is viewed at high resolution, as in inelastic electron
scattering, its wavefunction is seen to contain configurations where
its three ``valence" quarks are accompanied by further quarks and
antiquarks in its ``sea". The three quark configuration is thus merely
the simplest required to produce its overall positive charge and zero
strangeness. The question thus arises whether there are baryons for
which the minimal configuration cannot be satisfied by three quarks.

A baryon with positive amount of strangeness would be an example; the
positive strangeness requires an $\bar{s}$ and $qqqq$ are required for
the net baryon number, making what is known as a ``pentaquark" as the
minimal ``valence" configuration.

Hitherto unambiguous evidence for such states in the data has been
lacking; their absence having been explained by the ease with which
they would fall apart into a conventional baryon and a meson with
widths of many hundreds of MeV.  It is perhaps this feature that
creates the most tantalising challenge from the perspective of QCD:
why does $\Theta$ have width below 10MeV, perhaps no more than
1MeV\cite{1mev}.

If the data comprising the evidence as presented at this conference
are being correctly interpreted, they suggest that the $\Theta$ is
being produced with probability similar to that of the negative
strangeness $\Lambda(1520)$. This suggests that the $\Theta $ is
produced by the strong interaction between $KN$. However, such a
strength seems to be at odds with the implied feeble decay strength
implied by a 1MeV width into $KN$, unless perhaps $\Theta$ is produced
by the strong decay from some state $\Theta^*$, which is produced
strongly by KN and has width of $\geq O(100)$MeV. The other
possibility is that the production cross section of $\Theta$ is
$O(10-100)$ smaller than that of $\Lambda(1520)$ (there emerged some
hints after the conference that this might indeed be the
case\cite{saphir2})

However, it may be premature to seek radical solutions given the
nature of the current evidence\cite{leps}. The most immediate concern
must be to establish not simply the spin and parity of the $\Theta$,
or other examples like it, but to verify that it indeed exists and is
not some combination of statistical fluctuations, some complex novel
dynamical background effect that has been overlooked, or psychological
desire to be attracted by small positive signals while arguing away
any compensating negative results. In the immediate term, a dedicated
high statistics experiment involving photoproduction at Jefferson
Laboratory, planned to take data in 2004 may help to settle some of
these questions.

Whether or not it turns out to be real, the stimulus to theory has
already reinvigorated interest in the Skyrme model (which even
predicted that such a state should exist, at such a mass, though
admittedly not with a width so small) and the pentaquark dynamics of
the quark model. The Skyrme model and the quark model are both rooted
in QCD though their relation has been obscure. Considerable
theoretical attention into their relation has been stimulated by the
$\Theta$ studies. (Following the conference there has appeared a paper
which suggests\cite{itz} that the exotic $\Theta$ is an artifact of
the rigid rotator approach to the Skyrme model, and that in the
$SU(3)_F$ limit the \10bar does not form. )

Skyrme's model, when extended to incorporate strangeness, implied that
the lightest baryon families consisted of ({\bf 8}$\frac{1}{2}^+$)
which includes the nucleons, and a ({\bf 10}$\frac{3}{2}^+$) which
includes the $\Delta,\Omega^-$.  This far its predicted pattern is
like that of the quark model based on three quarks interacting with
QCD forces and also as seen in the data. However, it was noticed that
in this Skyrme model, there is a further family of ten (transforming
like a ``ten-bar", of SU(3)-flavour) with $J^P=\frac{1}{2}^+$. This is
the family that can not be formed from three quarks and requires the
pentaquark as a minimum configuration.

Initially it was thought that the pentaquark would lead to negative
parity for the lightest states, in contradiction to the Skyrme model
prediction of positive parity. However, the color magnetic forces of
QCD, when combined with constraints on flavor and spin required by
fundamental symmetries (such as Bose symmetry and the Pauli exclusion
principle) cause the lightest observable states plausibly to contain
one unit of internal angular momentum and thereby have positive parity
\cite{jw,kl1}.

However, there does appear to be a significant potential difference
between the models, which should be experimentally testable. Both
predict that there are two further exotic members of the ``ten-bar"
family: they have strangeness minus two, like the familiar $\Xi$
baryons, but whereas the familiar $\Xi$ states have electric charges 0
or -1, these can have 0,-1 and also +1 or -2. Positively charged or
doubly negatively charged baryons with strangeness minus two are
hitherto unknown.

And this is where the potential difference arises. In the formulation
of the Skyrme model for broken SU(3) in\cite{dpp}, the mass gap
between the $\Theta$ and these $\Xi$ has to be {\bf larger} than that
in the conventional ten, spanned by the $\Delta(1236)$ and
$\Omega^-$(1672). This appears to be unavoidable if the \10bar masses
are to be above those of the familiar decuplet. Indeed, they predicted
this gap in the ``ten-bar" to be some 540MeV leading to a mass for the
$\Xi$ exceeding 2GeV. In the pentaquark picture, by contrast, one need
only pay the price for one extra strange mass throughout the \10bar.
This implies a relatively light mass for the $\Xi$ $\sim 1700$MeV with
the possibility that these states also could be relatively stable.

I will now describe the wavefunctions of the pentaquark in more detail
to show that there is {\bf no} simple mapping onto the Skyrme model as
initially presented in \cite{dpp}.

\subsection*{\10bar Wavefunctions}

To get a feeling for a \10bar, first recall the most familiar decuplet
of baryons. This forms a large inverted triangle with the $\Omega^-$
at its pointed base and $\Delta^{++};\Delta^{-}$ at the two extremes
of its ``shoulders"; the strangeness spans 0 to -3. Now consider the
corresponding antibaryons, making a \10bar. Now we will have the
(anti- $\Omega)^+$ at the pointed head of the triangle and
(anti-$\Delta)^{--}$ and (anti-$\Delta)^{+}$ at the extremes of its
base; the strangeness spans +3 to 0. Note the electric charges of
these states. The \10bar of interest in the present story is like this
but with the magnitudes of strangeness being two units less throughout
than the antibaryon one just described. Thus instead of the
(anti-$\Omega)^+$ $(S=3)$ at the pointed head of the triangle we have
$\Theta^+$(S=1). In place of the (anti-$\Delta)^{--}$ and
(anti-$\Delta)^{+}$ $(S=0)$ at the extremes of its base we have the
exotic ($S=-2$) $\Xi^{--};\Xi^{+}$.

Thus we see the presence of three exotic correlations of strangeness
and charge. The $\Theta^+$ is what is claimed to have been discovered;
the $\Xi^{--};\Xi^{+}$ are a remaining challenge.

We all know how to write the wavefunctions for a \10bar made of three
antiquarks. However, there appears to be some confusion about the
analogous wavefunctions for a \10bar made of pentaquarks.  In
particular the form quoted in the discussion session here is not a
\10bar. Given this confusion I will describe here in a heuristic way,
how to build them. This will immediately expose essential differences
with the Skyrme model and suggest further novel implications in the
baryon spectrum.

I am going to view the $qqqq\bar{q}$ as two diquarks $qq$-$qq$
accompanying an antiquark.  To form the wavefunctions and take care of
their symmetries note first how the diquarks transform under
SU(3)$_f$. Define the antisymmetric diquark states cyclically under $u
\to d \to s$ so that (apart from normalisations)

\[
(ud) \equiv (ud-du) \to \bar{s}; (ds )\equiv (ds-sd)\bar{u};(su)
\equiv (su-us) \to \bar{d}
\]

Then take the traditional wavefunctions for antibaryons, retain one
antiquark and replace the others by the corresponding diquark.

The $\Theta$ state $(ud)^2\bar{s}$ is thus seen immediately to be
symmetric and analogous to the $\bar{\Omega}^+$. The analogues of the
$\bar{\Delta}^{--}$ and $\bar{\Delta}^+$ are then respectively
$(ds)^2\bar{u}$ and $(su)^2\bar{d}$. These form $\Xi$ states with
strangeness = -2 in our \10bar.

Before writing wavefunctions note immediately that there is only {\bf
  one} extra strange mass in the $\Xi$ states relative to the
$\Theta$. Thus in the pentaquark model one necessarily has low lying
exotic $\Xi$ states around 1700MeV if one identifies the
$\Theta(1540)$ to set the scale. This is different from the Skyrme
model as originally presented in \cite{dpp}.

This is an important fact worthy of some comment in view of the
prediction\cite{dpp} of the $\Theta$ in a \10bar in a version of the
Skyrme model.  However, it was critical in that prediction that the
mass gap from $\Theta$ to $\Xi$ is {\bf three} units of $\Delta(m_s -
m_d) \sim 150$MeV, as for the conventional (anti)decuplet of
(anti)$\Delta-(anti)\Omega$. In a pentaquark picture the mass gap is
only a single unit.

The difference comes from the way that\cite{dpp} implemented flavour
symmetry breaking.  A crucial assumption was that the SU(3) breaking
for $m_s \neq m_d$ depends linearly on the hypercharge Y=B+S such that
$M(Y) = M_0 -cY$ where $c>0$. For the familiar baryon {\bf 10} this is
equivalent to counting the number of strange quarks. However, this is
not a general axiom. It does not work for mesons, for example, where
$m(K^+) \equiv m(K^-)$ and $m(\omega) < m(\phi)$, nor for the octet
baryons where $m(\Sigma) > m(\Lambda)$. The origin of these masses are
immediately obvious in the quark model with hyperfine interactions.

The reason for the difference is that $s$ and $\bar{s}$ contribute
equally to the strange mass content, but cancel out in the
hypercharge. In the \10bar of interest here, the simple correspondence
familiar in the non-exotic {\bf 10} is lost. The mass gap from
shoulder to toe of the $10$, or from tip to base of the pyramid in the
\10bar is given by the difference in {\bf moduli} of the respective
strangeness. Thus for the familiar 10 or \10bar which run from
strangeness 0 to $\pm 3$ we have {\bf three} units of strange mass,
whereas for the case here which runs from strangeness +1 to -2 the
modular difference is only one.

Ref.\cite{dpp} forced the interval between the Theta and the N to be
1710-1540=170MeV and thereby inflate the mass splittings.  As we
already commented, the mass gap is $\frac{1}{3}m_s$ per stage in the
\10bar for the pentaquark whereas ref\cite{dpp} chose numbers with
more like one $m_s$ per unit gap.  Now their model at first sight
appears to hide beneath parameters $\alpha \beta \gamma$ (eq 16-18 in
hep-ph/9703373). However, this is not really so.  Critical is the mass
gap per unit of strangeness in table 1 of ref.\cite{dpp} which gives
the mass gaps per unit strangeness to be $1/8\alpha + \beta
-5/16\gamma$ for normal 10 and $1/8\alpha + \beta -1/16\gamma $ for
the novel \10bar. Hence in their convention where $1/8\alpha + \beta
<0$ then if $\gamma<0$ (see later) the mass gap per unit of
strangeness in their \10bar must be BIGGER than in the conventional
10.  The only way to get it smaller, as in the pentaquark picture
would be for $\gamma >0$.

So, what can one say about $\gamma$ in general?

First see eq 18 of ref\cite{dpp} and the comment at end of section 2:
``$I_1 > I_2$ so that the \10bar is heavier than the familiar
decuplet".  This tends to force $\gamma $ negative and toward $2
\beta/3$ (which is indeed in accord with their actual numbers of
$\gamma \sim $-107MeV and $\beta \sim$-156MeV in their eq 27).  So
there appears to be an inherent distinction between the Skyrme picture
of \cite{dpp} and the pentaquark, so long as m(\10bar) $>$ m({\bf
  10}).
 
There are other differences between the two pictures.  To motivate
these, we need first to look more carefully at the wavefunctions for
the other states in the multiplet.  The wavefunctions can be obtained
by applying the U-spin lowering operator to the $\Theta$. $U_-$
changes $d \to s$ or $\bar{s} \to -\bar{d}$.  $U_-$ commutes with the
Casimir operators of SU(3), and so under its operation one remains in
the \10bar.  Thus for example

\[
|p \rangle =
\frac{1}{2\sqrt{3}}\left([(ud-du)(su-us)+(su-us)(ud-du)]\bar{s}
  +(ud-du)^2\bar{d} \right)
\]
or more succinctly

\[
p = -\sqrt{\frac{2}{3}}\left[(ud)(su)_+ \right]\bar{s} -
\sqrt{\frac{1}{3}}(ud)^2\bar{d}
\]
We can expose the hidden \ss~ or \dd~ heuristically, though at the
expense of suppressing the above symmetries, by writing this in the
``shorthand" form

\[
p({\bf \bar{10}}) = uud\left( \sqrt{1/3} |d\bar{d}> +\sqrt{2/3}
  |s\bar{s}>\right).
\]
In similar fashion

\[
|\Sigma^+({\bf \bar{10}})> = U_- |p> \to uus\left(\sqrt{2/3}
  |d\bar{d}> + \sqrt{1/3} |s\bar{s}>\right)
\]

These are like familiar baryons with extra hidden strangeness or
hidden \dd~ in a specific weighted combination for the \10bar.  This
immediately allows one to count the total number of $s+\bar{s}$ in
each state.  In the N, for example, you get: $(1/3)\times (0) + (2/3)
\times 2 = 4/3$.  For the $\Sigma$: $(2/3) \times 1 + (1/3) \times 3 =
5/3$. Hence one sees explicitly the equal mass rule but with $m_s/3$
per unit change of strangeness, consistent with our earlier
observation that the total mass interval between the $\Theta$ and the
$\Xi^+ \equiv -(us)^2\bar{d}$ feels only {\bf one} extra strange
contribution.

Now we come to the interesting features, namely those states that are
not at the corners of the \10bar. These can also form octet
representations, whose wavefunctions are orthogonal to the above; they
are

\bea
p({\bf 8}) \to uud\left( \sqrt{2/3} |d\bar{d}> -\sqrt{1/3} |s\bar{s}>\right) \\
\Sigma^+({\bf 8}) \to uus\left(\sqrt{1/3} |d\bar{d}> - \sqrt{2/3}
  |s\bar{s}>\right) \eea

Counting the number of $s+\bar{s}$ one gets for the relative strange
mass content to the mass pattern in the octet $N:\Sigma:\Xi =
2/3:7/3:2$.

Photoproduction of the \10bar is interesting since the photon has
$U=0$ and so cannot cause transition from $p({\bf 8}) (U=1/2)$ to
$p$(\10bar)$(U=3/2)$. By contrast, the neutron is in a $U=1$ multiplet
for both {\bf 8} and \10bar, and hence $\gamma n({\bf 8}) \to$\10bar
is allowed.  To see this with the above wavefunctions, let the photon
convert to a \qq~ with amplitude proportional to the charge $e_q$;
form the transition amplitude by isolating the terms in the \10bar
wavefunction where $(q_iq_j)(q_kq_l)\bar{q}_l$ occur with the $q$ and
$\bar{q}$ of the same flavour adjacent to one another.  Thus

\[
p \to (ud-du)u \left[s\bar{s} - d\bar{d} \right]
\]
exposes the coupling to the mixed-antisymmetric conventional octet
proton uud state, and a $U=1$ state.  This explicitly shows that
$p({\bf 8}) \to p$(\10bar) transforms as $\Delta U =1$ whereby
photoproduction is forbidden.

The analogous exercise for a neutron gives

\[
n \to (ud-du)d \left[s\bar{s} - u\bar{u} \right]
\]
where the \qq~ piece now transforms as $V=1$, or equivalently as a
linear superposition of $I=1$ and $U=0$. The latter therefore allows
$\gamma n({\bf 8}) \to$\10bar.

Thus photoproduction could be imagined as a way to distinguish whether
the pentaquark $p^*$ is in {\bf 8} or \10bar.  However, it is at this
point, if not already, that one realises that the language of \10bar
and {\bf 8} is not really suitable. The symmetry breaking allows
mixing between the two multiplets and depending on the dynamics this
may tend toward the extreme which respectively maximises and minimises
the net $s + \bar{s}$ content. Thus the mass eigenstates may be
expected to tend toward the following (subscripts L and H for light
and heavy):

\[
N_L = (ud)^2 \bar{d}; N_H = (ud)(us)\bar{s}
\]

\noindent and

\[
\Sigma_L = (ud)(ds) \bar{u}; \Sigma_H = (ds)^2\bar{s}
\]

In this case we see that for the set of ``light" states, there is an
increase of order $m_s$ per strange gap, while the same is true for
their heavy counterparts until the final stage where the $\Xi$ is
lighter than the $\Sigma$, thereby preserving the ubiquitous rule that
there is only one unit of ``extra" strange mass between $\Theta$ and
$\Xi$.

Thus if one identifies the $m(\Theta) \sim 1540$MeV, one might
identify $m(N_H) \sim 1710$ (contrast the Skyrme model which
identified the \10bar with this state) and then have the prediction of
a lighter state, perhaps $m(N_L) \sim 1400$MeV, which could be related
to the Roper resonance\cite{jw}.

In the Skyrme model the \10bar has $J^P = \frac{1}{2}^+$; there is no
accompanying octet, and hence no possibility of mixing. In the
pentaquark model one might naively expect that the lowest lying states
are $J^P = \frac{1}{2}^-$; however, when the interquark QCD spin
dependent forces are taken into account one finds\cite{jw,kl1} that
octet and \10bar emerge lightest with $J^P = \frac{1}{2}^+$. However,
one also finds that they are partnered by $J^P = \frac{3}{2}^+$
multiplets too. Let's now look into this and assess experimental
tests.

\subsubsection*{Diquark Cluster Models}

Early evidence that mesons and baryons are made of the same quarks was
provided by the remarkable successes of the Sakharov-Zeldovich
constituent quark model, in which static properties and low lying
excitations of both mesons and baryons are described as simple
composites of asymptotically free quasiparticles with a flavor
dependent linear mass term and hyperfine interaction,

\begin{equation} 
M = \sum_i m_i + \sum_{i>j}  {{\vec{\sigma}_i\cdot\vec{\sigma}_j}\over{m_i\cdot
m_j}}\cdot v^{hyp} 
\end{equation} 
where $m_i$ is the effective mass of quark $i$, $\vec{\sigma}_i$ is a
quark spin operator and $v^{hyp}_{ij}$ is a hyperfine interaction.
 
As first pointed out by Karliner and Lipkin\cite{kl1}, a
single-cluster description of the ($uudd\bar s$) system fails because
of the repulsive interaction between the pairs of the same flavor,
which prevents binding. This leads one to consider dynamical
clustering into subsystems of diquarks or/and triquarks, which amplify
the attractive color-magnetic forces. There are two routes that emerge
naturally; one is that of\cite{kl1}, the other of Jaffe and
Wilczek\cite{jw}.  These naturally lead to $J^P=\frac{1}{2}^+$ as the
lowest mass states.

The first step is common and is based on the strong chromomagnetic
attraction between a $u$ and $d$ flavour when the $ud$ diquark is in
the $\bar {\hbox{\bf 3}}$ of the color $SU(3)$ and in the $\bar
{\hbox{\bf 3}}$ of the flavor $SU(3)$ and has $I=0, S=0$, like the
$ud$ diquark in the $\Lambda$.

Such an idea has a long history, being the source of the $\Lambda -
\Sigma$ mass difference, a possible linkage with the dominance of $u(x
\to 1)$ in deep inelastic structure functions and of the maximisation
of the polarisation asymmetry in this same limit. Such attraction
between quarks in the color $\bar{3}$ channel halves their effective
charge, reduces the associated field energy and is a basis of color
superconductivity in dense quark matter\cite{wilc}. There is the
implied assumption that such a ``diquark" may be compact, an effective
boson ``constituent", which is hard to break-up and hard for its
constituents to rearrange with other quarks or antiquarks in the bound
state.  I shall refer to this by $[(ud)_0]$, the subscript denoting
its spin, and the [ ] denoting the compact quasiparticle.
 
JW consider the following subcluster for the pentaquark:
$[(ud)_0][(ud)_0]\bar{s}$. KL also start with the $[(ud)_0]$ seed, but
regard the remainder as a strongly bound ``triquark"
$[(ud)_1\bar{s}]$.  This internal structure is chosen to give the
minimum energy to the triquark system; the $(ud)_1$ is coupled to spin
1, colour $6$; the $\bar{s}$ couples to the $u$ or the $d$ to net spin
0. Thus we see there is this difference in details between the two
approaches. First I will describe their dynamics and see what
consequences there are.

For JW the two $(ud)$ must combine to make $3_c$ in order to
neutralise the $\bar{s} = \bar{3}$; since $\bar{3} \times \bar{3} \to
3$ is antisymmetric in colour, and since the $(ud)_0$ boson pair must
be symmetric overall this implies that they are in P-wave (spatially
antisymmetric).This gives a negative parity that combines with the
negative parity of $\bar{s}$ to give an overall positive parity
system. Thus one has $J^P =\frac{1}{2}^+;\frac{3}{2}^+$ pentaquark
systems. It is possible to identify the mass with the $\Theta$ (see
later); the metastability can be accommodated by insisting that the
quasiparticles in $[(ud)_0][(ud)_0]\bar{s}$ prevent simple
rearrangement to overlap with $[(ddu)][u\bar{s}]$, which are the $NK$
colour singlet hadrons.

Whereas JW take the other $ud$ diquark also to be in this
configuration and then put the two diquarks in relative P-wave (by
Bose symmetry after the colour is taken account of), KL by contrast
took the remaining $ud\bar{s}$ and looked for the configuration in
color and spin which would optimize the total (five-body) hyperfine
interaction.

Karliner and Lipkin divide the system into two color non-singlet
clusters which separate the pairs of identical flavor.  The two
clusters, a $ud$ diquark and a $ud\bar s$ triquark, are separated by a
distance larger than the range of the color-magnetic force and are
kept together by the color electric force. Therefore the color
hyperfine interaction operates only within each cluster, but is not
felt between the clusters. They associate the $[(ud)_0]$ with the
non-strange piece of the $\Lambda(1110)$ baryon.

Within the $[(ud)_1\bar{s}]$ the strange subsystems $u\bar{s}$ and
$d\bar{s}$ are assumed to be in spin 0, by colour-spin forces
analogous to the way that the K is lighter than $K^*$. If the diquark
and triquark are in relative S-wave, then the colour attractions act
among all the constituents leading to a freeze out that would be a
$KN$ S-wave system. In P-wave, the two separate quasi-particles avoid
the contact hyperfine forces (their generalisations to the Fermi-Breit
effects are not discussed).  These identifications help them to argue
that the mass of the system agrees with that of the $\Theta$.
Analogous to JW, the metastability can be accommodated by insisting
that the quasiparticles in $[(ud)_0][(ud)_1\bar{s}]$ prevent simple
rearrangement to overlap with $[(ddu)][u\bar{s}]$, which are the $NK$
colour singlet hadrons.

KL consider also heavy analogues $[(ud)_0][(ud)_1\bar{Q}]$ where $Q
\equiv s;c;b$. They do not consider $Q \equiv u,d$, possibly because
this enables annihilation with the like-flavoured quark in the
triquark, thereby destroying the stability and overlapping with
conventional $udu$ or $udd$ baryons.  However, we shall see that such
states can have non-trivial consequences.
 
JW do consider $[(ud)_0][(ud)_0]\bar{Q}$ with $Q \equiv u,d$. The
dynamical difference is that the quasi-particle nature of the separate
$([(ud)_0])$ may suppress the annihilation with the like flavour,
enabling the states $[(ud)_0][(ud)_0]\bar{Q}$ with $Q \equiv u,d$ to
have an existence. Such states would be expected to lie
$O(100-150)MeV$ below the $\Theta$, and they identify them with the
Roper $n;p(1440)$ nucleon resonances.
 
The mass of the exotic $\Xi^+$ depends rather sensitively on the
effects of clustering. First one needs the effective mass of the
diquarks. The mass difference $m(\Delta) - m(N)$ implies that
 
 \[
 \Delta m[(ud)_0] = -150 MeV; \Delta m[(ud)_1] = +50 MeV
 \]
 relative to their mean masses. The strange counterparts follow from
 $m(\Sigma^*) - m(\Sigma)$ and implies
 
 \[
 \Delta m[(us)_0] = -96 MeV.
 \]
 Thus
 
 \[
 \Delta m(\Xi^+ - \Theta^+) = \Delta m(\bar{s} - \bar{d}) + 2 \Delta
 m( [(us)_0] - [(ud)_0]) \sim 230-250MeV.
 \]
 So 1750-1800MeV is the mass range one obtains with this level of
 approximation, which is still significantly below the original
 prediction in ref\cite{dpp}. There is further uncertainty in
 estimating the mass in that the orbital excitation of the $[us][us]$
 and $[ud][ud]$ will have different energetics and the $\vec{L} \cdot
 \vec{S}$ shifts are also dependent on the flavours. These could
 easily add further uncertainties of $\pm 50$MeV.  If and when the
 $\Xi^{+(--)}$ are discovered, along with their $J^P = \frac{3}{2}^+$
 counterparts, their masses will enable the systematics of the
 clustering to be determined by fitting the above.
 
 \subsection*{Other pentaquark states}
 
 At first sight the narrowness of the $\Theta$ would seem to argue
 against any identification of the broad Roper resonance as a nucleon
 analogue. However, this need not be so. Some of the following
 thoughts emerged from discussion with Maltman at the
 conference\cite{maltman}.
 
 For a simple attractive square well potential of range 1fm. the width
 of a P-wave resonance 100MeV above $KN$ threshold is of order
 200MeV\cite{jw,maltman}. However, this has not yet taken into account
 any price for recoupling colour and flavour-spin to overlap the
 $(ud)(ud)\bar{s}$ onto colour singlets $uud$ and $d\bar{s}$ say for
 the $KN$. In amplitude, starting with the Jaffe-Wilczek
 configuration, the colour recoupling costs $\frac{1}{\sqrt{3}}$ and
 the flavour-spin to any particular channel (e.g. $K^+n$) costs a
 further $\frac{1}{4}$. This appears to be akin to the factor
 $1/2\sqrt{6}$ found in ref\cite{maltman} for the isospin-spin-colour
 overlap in their conventions. They go further and consider the mixing
 with the Karliner-Lipkin configuration gives for the lowest
 eigenstate a suppression of $\frac{1}{25}$\cite{maltman}. Hence a
 width of $O(1-10 MeV)$ for $\Theta \to KN$ may be reasonable.

 The decay involves tunneling through the P-wave barrier, which by
 analogy with $\alpha$-decay is exponentially sensitive to the
 difference between the barrier height and the kinetic energy of the
 state. This can affect the width of a spin $\frac{3}{2}$
 ``$\Theta^*$" partner, whose mass $> m(\Theta)$, arising from the
 $\vec L \cdot \vec S$ splitting effects in the pentaquark system.
 This splitting is model dependent\cite{dudek} but might be as small
 as $\sim 30-80$MeV.  Given the exponential sensitivity in a tunneling
 width, the $\Theta^*$ could be high enough in the well to be broad.
 In any event such a state should be sought. If the above is a guide,
 only $\Theta^* \to \Theta \gamma$ is kinematically allowed as a
 transition.  If its mass exceeds 1820MeV it becomes possible for a
 strong decay width $\Theta^* \to \Theta \pi\pi$ to feed at least some
 of the $\Theta$ signal.

 The exponential sensitivity could frustrate attempts to differentiate
 between the pentaquarks and (original) Skyrme model in connection
 with the exotic $\Xi$ states. In Skyrme these are above 2GeV and
 relatively broad; in pentaquarks they are $\sim 1700$MeV and hence
 the possibility of being relatively narrow, perhaps only 50\% broader
 than the $\Theta$\cite{jw}. However, exponential dependence could
 cause even 1700 MeV to be high enough in the well to give a broad
 width to $\Xi \pi$. It would be galling if the $\Theta$ were the only
 sharp state. (See also ref.\cite{itz} and comments earlier about the
 potential non-binding of the \10bar in the Skyrme picture.)
 
 For the Roper, the naive mass of the $uudd\bar{d}$, by comparison
 with the $\Theta$, would be $\sim 1400$MeV. Its physical mass of
 1430-1470MeV\cite{PDG} may therefore have it elevated in the
 potential, where the exponential behaviour drives its large width.
 These possibilities need more careful study in models.
 
 However, for the non-exotic states the picture is not so simple. As
 stressed earlier, there is no absolute meaning to a pentaquark
 configuration when a $qqq$ constituent state can carry the same
 overall quantum numbers. Thus the $uudd\bar{d}$ is at best the
 $SU(3)$ flavour state within the pentaquark wavefunction of the
 Roper, or for that matter, of the nucleon. Mass eigenstates will be
 mixtures of $qqq$, these pentaquark states and higher configurations.
 The nucleon may be $qqq$ in leading order with its pentaquark
 components, which naively exist at higher mass scale, revealed with
 increasing $q^2$. For the Roper, the mass scales of the pentaquark
 and the excited $qqq$ components may compete. Certainly both states
 require $qqq$ presence in order to understand the -2:3 relative
 amplitudes for $\gamma n:\gamma p$ magnetic moment/transitions.
 Furthermore, the existence of the $\Delta(1660)$ as a potential
 partner of the Roper (analogous to the $\Delta(1230)$ for the
 nucleon) plays an essential role in disentangling these states.
 Improved data on its photo-excitation from Jefferson Laboratory could
 help here. Note there is no simple place for such a state if the
 Roper were pure pentaquark.
 
 \subsection*{The Nucleon Sea}
 
 As stressed above: the number of quarks in a nucleon is not a
 meaningful quantity. As $q^2$ varies the nucleon's structure is
 probed on ever finer resolution and the sea of \qq~ is exposed.  All
 that we can say is that three is the minimum number of quarks
 required to satisfy the overall quantum numbers. Thus there are
 pentaquark, heptaquark and ad-inf.-quark components to the
 wavefunction of any state which transforms like a nucleon.
 
 In the case of a positively charged positive strangeness baryon, the
 $uudd\bar{s}$ is the minimal configuration compatible. Thus for such
 a state there is meaning to the pentaquark as the minimum ``valence"
 wavefunction. Within the pentaquark sector, which contains this
 state, there are configurations which transform like a nucleon and
 these are in general mixtures of octet and \10bar, as described above

\[
uud \left[cos\theta (d\bar{d}) + sin \theta(s\bar{s}) \right]
\]

The QCD forces that have led to the $\Theta$ being the lightest
pentaquark state will lead to the above as the lightest analogues with
nucleon quantum numbers. Note in particular that it is the attractive
forces between $ud$ pairs that have favoured these and contrast this
with the $uudu\bar{u}$ pentaquark, which is in a {\bf 27} of flavour
SU(3), and is pushed to higher energy by the repulsive forces between
the symmetric $u$ quarks.
 
If indeed the lightest proton excitation is the $(ud)^2 \bar{d}$, then
this would be a natural candidate for the leading piece of the five
body proton Fock state. This is of course what the folklore for
failure of Gottfried sum rule requires; the asymmetry that leads to
the "missing" $(ud)uu\bar{u}$ in this picture is because the 27 is
pushed up relative to the \10bar-{\bf 8} mixture.  The antiquark sea
is also naively polarised "against the flow" too.
 
There is an interesting duality between this and other interpretations
of this flavour asymmetry in the sea. One is to invoke Pauli blocking
of the \uu~ due to the extra $u$ flavour in the valence proton. In
effect, this is an essential feature subsumed within the arguments of
\cite{jw,kl1}, which addressed the pentaquark configuration for the
$\Theta$.
 
Another approach has been to consider the $N\pi$ cloud of the nucleon.
The essence is that $p \to n \pi^+(u\bar{d})$ feeds the $\bar{d}$
whereas the $\bar{u}$ is energetically disfavoured, requiring $p \to
\Delta^{++} \pi^-(d\bar{u})$. The QCD forces that push the $m(\Delta)
> m(n)$ are the same that distorted the pentaquark configurations,
favouring the {\bf 8}-\10bar over the {\bf 27}. So the role of
multiquark configurations, and their mapping onto the meson-baryon
sectors, are all pervading. It is a question of approximation as to
whether one or other dominates, or whether both play competing roles.
Whether the $\Theta(1540)$ will turn out to be the first real evidence
for a state with a minimal pentaquark ``valence" configuration, or
merely a strange story to tell future generations, it has certainly
raised challenging questions and is leading to some unexpected
insights.

 \subsection*{Postscript on $\Theta$ for future historians}

 A vote at the conference on whether $\Theta$ can be regarded as (i)
 an established resonance, (ii) jury still out, (iii) is not a
 resonance, was split approximately $\frac{1}{4}; \frac{1}{2}$ and
 $\frac{1}{4}$ respectively.  The total number of votes cast was
 $O(100)$.  People who were involved in any of the relevant
 experiments or had written theory papers on $\Theta$ were excluded
 from the vote. There appeared to be a slight tendency for senior
 experimentalists to vote in category (iii). Whether this is because
 of experience with the machinations of statistics in the past, or
 frustration at having overlooked a major discovery, is for
 psychologists to debate.

 \bc {\bf Acknowledgments} \ec
 
 I am deeply indebted to Mrs Hiltscher and Bernd Lewandowski for their
 help in preparing this talk and congratulate the organisers for a
 superb conference. In addition to talking with many people at the
 conference, I have profited from discussions about pentaquarks and
 exotic hadrons with J.Dudek, R.Jaffe, M.Karliner and H.Lipkin.

\end{document}